\pdfoutput=1
\documentclass{article}

\usepackage{amssymb,amsfonts,amsmath}
\usepackage{cite,enumerate,float}
\usepackage{color}
\usepackage{tikz}
\usetikzlibrary{arrows,snakes,backgrounds}
\usepackage{hyperref}

\usepackage{amssymb,amsfonts,amsmath,stmaryrd}
\usepackage{cite,enumerate,float,indentfirst}
\usepackage{color}
\usepackage{tikz}
\usetikzlibrary{arrows,snakes,backgrounds,calc}
\usepackage{ytableau}
\usepackage[vcentermath]{youngtab}
\usepackage{pict2e}

\def\be{\begin{eqnarray}}
\def\ee{\end{eqnarray}}
\def\nn{\nonumber}

\def\be{\begin{eqnarray}}
\def\ee{\end{eqnarray}}
\def\nn{\nonumber}

\def\p{\partial}

\def\Tr{{\rm Tr}\,}

\definecolor{red}{rgb}{1,0,0}
\definecolor{orange}{rgb}{1,0.5,0}
\definecolor{violet}{rgb}{0.7,0,1}

\hoffset= - 1.0in         

\begin{document}

\title{\vspace{1.5cm}\bf
$W_{1+\infty}$ and $\widetilde W$ algebras, and Ward identities
}

\author{
Ya. Drachov$^{a,}$\footnote{drachov.yai@phystech.edu},
A. Mironov$^{b,c,d,}$\footnote{mironov@lpi.ru,mironov@itep.ru},
A. Popolitov$^{a,c,d,}$\footnote{popolit@gmail.com}
}

\date{ }

\maketitle

\vspace{-6.5cm}

\begin{center}
\hfill FIAN/TD-16/23\\
\hfill IITP/TH-22/23\\
\hfill ITEP/TH-28/23\\
\hfill MIPT/TH-21/23
\end{center}

\vspace{4.5cm}

\begin{center}
$^a$ {\small {\it MIPT, Dolgoprudny, 141701, Russia}}\\
$^b$ {\small {\it Lebedev Physics Institute, Moscow 119991, Russia}}\\
$^c$ {\small {\it NRC ``Kurchatov Institute", 123182, Moscow, Russia}}\\
$^d$ {\small {\it Institute for Information Transmission Problems, Moscow 127994, Russia}}
\end{center}

\vspace{.1cm}

\begin{abstract}
It was demonstrated recently that the $W_{1+\infty}$ algebra contains commutative subalgebras associated with all integer slope rays (including the vertical one). In this paper, we realize that every element of such a ray is associated with a generalized $\widetilde W$ algebra. In particular, the simplest commutative subalgebra associated with the rational Calogero Hamiltonians is associated with the $\widetilde W$ algebras studied earlier. We suggest a definition of the generalized $\widetilde W$ algebra as differential operators in variables $p_k$ basing on the matrix realization of the $W_{1+\infty}$ algebra, and also suggest an unambiguous recursive definition, which, however, involves more elements of the $W_{1+\infty}$ algebra than is contained in its commutative subalgebras. The positive integer rays are associated with $\widetilde W$ algebras that form sets of Ward identities for the WLZZ matrix models, while the vertical ray associated with the trigonometric Calogero-Sutherland model describes the hypergeometric $\tau$-functions corresponding to the completed cycles.
\end{abstract}

\bigskip

\section{Introduction}

$\widetilde W$ algebras were first constructed many years ago as algebras of constraints in two-matrix models \cite{Gava,AS}. Later, they also emerged in the character phase of generalized Kontsevich model \cite{GKMU}, where it was noted that, in fact, there are two series of the $\widetilde W$ algebras: $\widetilde W^{(\pm,n)}$. The simplest $\widetilde W^{(\pm,n)}$ algebra is nothing but the Borel subalgebra of the Virasoro algebra. Higher spin algebras are no longer Lie algebras, and can be described by commutation relations \cite{Gava,GKMU}. However, more convenient is to use their representation in terms of an infinite set of variables $p_k$: then, the algebra is given by manifest expressions for its generators as graded differential operators in $p_k$. These expressions can be obtained from defining recurrent relations, or from a realization of the generators in terms of a matrix $\Lambda$ such that $p_k=\Tr\Lambda^k$.

Though the $\widetilde W$ algebras emerged later in various contexts related to matrix models (see, e.g., \cite{AMMN2,MMsc}), their meaning remained unclear. In the present paper, we make a step in revealing their meaning and demonstrate that the $\widetilde W$ algebras are naturally associated with the $W_{1+\infty}$ algebra  \cite{Pope,FKN2,BK,BKK,KR1,FKRN,Awata,KR2,Miki}, and with WLZZ models \cite{China1,China2}. In fact, this relation was already preliminary discussed in \cite{MMsc}.

This relation of the $\widetilde W$ algebras with the $W_{1+\infty}$ algebra is as follows: as it was demonstrated in \cite{MMMP1}, the $W_{1+\infty}$ algebra contains infinitely many commutative families, which are called {\it integer rays}, {\it rational rays} and {\it cones}. These names are related to the $2d$ integer lattice of generators of the $W_{1+\infty}$ algebra: the vertical axis describes the maximal spin of generator, and the horizontal one, the grading. The integer rays drawn in Fig.1 are just the rays with integer slopes. Rays with rational slopes are called rational rays, and a unification of rays is called cones \cite{MMMP1}.

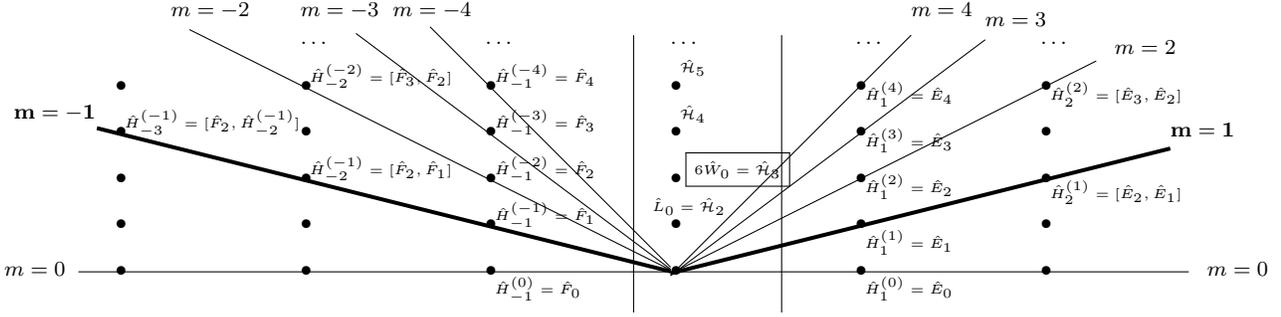
\begin{figure}[h]
\setlength{\unitlength}{.7pt}
\begin{picture}(300,170)(-350,-15)
{\footnotesize
\put(-20,-20){\line(0,1){150}}
\put(60,-20){\line(0,1){150}}

\put(0,0){\mbox{$\bullet$}}
\put(0,25){\mbox{$\bullet$}}
\put(0,50){\mbox{$\bullet$}}
\put(0,75){\mbox{$\bullet$}}
\put(0,100){\mbox{$\bullet$}}
\put(0,125){\mbox{$\ldots$}}
\put(-10,35){\mbox{\tiny$\hat L_0=\hat {\cal H}_2$}}
\put(8,55){\mbox{\tiny $\boxed{6\hat W_0=\hat {\cal H}_3}$}}
\put(5,85){\mbox{\tiny$\hat {\cal H}_4$}}
\put(5,110){\mbox{\tiny$\hat {\cal H}_5$}}

\put(-100,0){
\put(0,0){\mbox{$\bullet$}}
\put(0,25){\mbox{$\bullet$}}
\put(0,50){\mbox{$\bullet$}}
\put(0,75){\mbox{$\bullet$}}
\put(0,100){\mbox{$\bullet$}}
\put(0,125){\mbox{$\ldots$}}
\put(5,-10){\mbox{\tiny $\hat H_{-1}^{(0)}=\hat F_0 $}}
\put(5,30){\mbox{\tiny$\hat H_{-1}^{(-1)}=\hat F_1
$}}
\put(5,55){\mbox{\tiny$\hat H_{-1}^{(-2)}=\hat F_2$}}
\put(5,80){\mbox{\tiny$\hat H_{-1}^{(-3)}=\hat F_3$}}
\put(5,105){\mbox{\tiny$\hat H_{-1}^{(-4)}=\hat F_4$}}
}

\put(-200,0){
\put(0,0){\mbox{$\bullet$}}
\put(0,25){\mbox{$\bullet$}}
\put(0,50){\mbox{$\bullet$}}
\put(0,75){\mbox{$\bullet$}}
\put(0,100){\mbox{$\bullet$}}
\put(0,125){\mbox{$\ldots$}}
\put(5,55){\mbox{\tiny $\hat H_{-2}^{(-1)}=[\hat F_2,\hat F_1]$}}
\put(5,105){\mbox{\tiny $\hat H_{-2}^{(-2)}=[\hat F_3, \hat F_2]$}}
}

\put(-300,0){
\put(0,0){\mbox{$\bullet$}}
\put(0,25){\mbox{$\bullet$}}
\put(0,50){\mbox{$\bullet$}}
\put(0,75){\mbox{$\bullet$}}
\put(0,100){\mbox{$\bullet$}}
\put(5,80){\mbox{\tiny $\hat H_{-3}^{(-1)}=[\hat F_2,\hat H_{-2}^{(-1)}]$}}
}

\put(100,0){
\put(0,0){\mbox{$\bullet$}}
\put(0,25){\mbox{$\bullet$}}
\put(0,50){\mbox{$\bullet$}}
\put(0,75){\mbox{$\bullet$}}
\put(0,100){\mbox{$\bullet$}}
\put(0,125){\mbox{$\ldots$}}
\put(5,-10){\mbox{\tiny$\hat H_1^{(0)}=\hat E_0$}}
\put(5,15){\mbox{\tiny$\hat H_1^{(1)}=\hat E_1$
}}
\put(5,45){\mbox{\tiny$\hat H_1^{(2)}=\hat E_2$}}
\put(5,70){\mbox{\tiny$\hat H_1^{(3)}=\hat E_3$}}
\put(5,95){\mbox{\tiny$\hat H_1^{(4)}=\hat E_4$}}
}

\put(200,0){
\put(0,0){\mbox{$\bullet$}}
\put(0,25){\mbox{$\bullet$}}
\put(0,50){\mbox{$\bullet$}}
\put(0,75){\mbox{$\bullet$}}
\put(0,100){\mbox{$\bullet$}}
\put(0,125){\mbox{$\ldots$}}
\put(5,42){\mbox{\tiny $\hat H_2^{(1)}=[\hat E_2,\hat E_1]$}}
\put(5,95){\mbox{\tiny $\hat H_2^{(2)}=[\hat E_3, \hat E_2]$}}
}

\put(0,2){\line(-4,1){310}}
\put(0,1){\line(4,1){270}}
\put(0,3){\line(-2,1){260}}
\put(0,1){\line(2,1){230}}
\put(0,4.5){\line(-1,1){130}}
\put(0,0){\line(1,1){130}}
\put(0,0){\line(4,3){170}}
\put(0,3.4){\line(-4,3){170}}


\put(130,140){\mbox{$m=4$}}
\put(170,135){\mbox{$m=3$}}
\put(240,120){\mbox{$m=2$}}
\put(270,75){\mbox{${\bf m=1}$}}
\put(-150,140){\mbox{$m=-4$}}
\put(-200,140){\mbox{$m=-3$}}
\put(-270,140){\mbox{$m=-2$}}
\put(-355,85){\mbox{${\bf m=-1}$}}

\put(-320,2){\line(1,0){600}}
\put(-360,0){\mbox{$m=0$}}
\put(290,0){\mbox{$m=0$}}



\linethickness{1.5pt}
\put(0,2.2){\line(-4,1){310}}
\put(0,1.2){\line(4,1){270}}

}
\end{picture}
\caption{Commutative families (integer rays) on the $2d$ integer lattice of generators of the $W_{1+\infty}$ algebra.}
\label{fig:plane}
\end{figure}

The commuting elements of the algebra lying on these rays are constructed in the following way: one constructs two sets of generating elements, each of them being the first Hamiltonian of the corresponding commutative family:
\be
	\hat E_n = {\rm ad}_{\hat W_0}^n \hat E_0, \quad  \hat F_n = {\rm ad}_{\hat W_0}^n \hat F_0\,.
\ee
and all other commutative Hamiltonians at any fixed $m$ are generated as
\be
 {\hat H}_k^{(m)} = {\rm ad}_{\hat E_{m+1}}^{k-1} \hat E_m \, , \quad \text{and} \quad  {\hat H}_{-k}^{(-m)} = {\rm ad}_{\hat F_{m+1}}^{k-1} \hat F_m
 \ee
There is also the commutative vertical ray: the set of zero grading elements generated as
\be
 [\hat F_i,\hat E_j]=\hat{\cal H}_{i+j}
\ee
Thus, all the elements are generated by repeated commutators of the three elements: $\hat W_0$, $\hat E_0$ and $\hat F_0$. The $W_{1+\infty}$ algebra can be also realized in terms of graded differential operators in variables $p_k$ \cite{MMMP1} so that these three elements are
\be
\hat W_0=\frac{1}{2}\sum_{a,b=1} \left(abp_{a+b}\frac{\p^2}{\p p_a\p p_b} + (a+b)p_ap_b\frac{\p}{\p p_{a+b}}\right)
		+ N\sum_{a=1} ap_a\frac{\p}{\p p_a}+{N^3\over 6},\ \ \ \ \ \ \ \hat E_0=p_1,\ \ \ \ \ \ \ \hat F_0={\p\over\p p_1}
\ee
Then, as follows from the grading, one can represent all $H_k^{(m)}$ in the form
\be
H_n^{(m)}=\sum_kp_k{\cal O}^{(m,n)}_{k-n}
\ee
As a matter of fact, there is a similar representation for the $H_{-k}^{(-m)}$ families:
\be
H_{-n}^{(-m)}=\sum_kp_k{\cal O}^{(-m,-n)}_{k+n}
\ee
though it does not follow from general arguments. Moreover, this is no longer the case for the rational rays.

Such representations of the commutative families was first noted in \cite{MMsc} for $m=1$ families, where it was pointed out that ${\cal O}^{(\pm 1,n)}$ are nothing but the elements of the $\widetilde W^{(\mp,n)}$ algebras. Here we extend this observation to all commutative families associated with the integer rays.

An important feature of the $\widetilde W^{(m,n)}$ algebras is that they realize sets of the Ward identities satisfied by partition functions $Z_k^{(m)}$ of the WLZZ models:
\be
Z_n^{(m)}=e^{{1\over n}H_n^{(m)}}\cdot 1=\sum_R
\left({S_R\{p_k=N\}\over S_R\{p_k=\delta_{k,1}\}}\right)^mS_R\{p_k=\delta_{k,n}\}S_R\{p_k\}
\ee
where $S_R\{p_k\}$ denotes the Schur function as a function of power sums $p_k$, and $R$ is a partition. These models at $m=1$ admit a two-matrix model representation, and, in this case, $\widetilde W^{(1,n)}$ is the standard $\widetilde W^{(-,n)}$ algebra, and it is known \cite{AMMN2} to form a set of Ward identities for this two-matrix model.

The paper is organized as follows. In section 2, we describe the standard $\widetilde W$ algebras. In section 3, we introduce the extension of $\widetilde W$ algebras related to the $W_{1+\infty}$ algebra, and, in section 4, describe their definition using the recursive relations. This requires introducing more elements of the $W_{1+\infty}$ algebras, not related to the commutative families. In section 5, we concentrate on the $\widetilde W$ algebra associated with the vertical ray. In section 6, we discuss the Ward identities in one of the branches of the WLZZ series of models and explain that the generalized $\widetilde W$ algebras describe the Ward identities (constraint algebra) in these models. Section 7 contains some discussion and concluding remarks.

\section{Standard $\widetilde W$ algebras}

In this section, we describe the standard $\widetilde W$ algebras \cite{Gava,GKMU,UFN}. We also introduce an $N\times N$ matrix $\Lambda$ such that its traces are $p_k = \Tr\Lambda^k$. We imply that $N$ is large (formally, one has to bring $N$ to $\infty$, see \cite{V} for an accurate description of this procedure). Then, generators of the $\widetilde W$ algebras are defined from
\be
\left.
\left(\frac{\partial}{\partial \Lambda}\right)^{n} f(p) =
\sum_{k=1} \Lambda^{k}\widetilde { W}_{k + n}^{(+,n)}(p)
f(p)
\right|_{{p_a =  { \Tr} \Lambda^{a}}}
\label{W-tilde-1}
\ee
and\footnote{This formula can be also rewritten in the form
$$
\left.
\left(-\left(\det\Lambda\right)^{-N}\frac{\partial}{\partial \Lambda^{-1}}\left(\det\Lambda\right)^{N}\right)^{n} f(p) =
\sum_{k=1} \Lambda^{-k}\widetilde { W}_{k - n}^{(-,n)}(p)
f(p)
\right|_{{p_a =  { \Tr} \Lambda^{-a}}}
$$
in accordance with the general principle of turning from the left hand side of Fig.1 to the right hand side, when the operators ${\cal O}(\Lambda)$ go to operators: $-\det^{-N}\Lambda\cdot {\cal O}(\Lambda^{-1})\cdot\det^N\Lambda$ \cite{MMCal,MMMP1}.
}
\be
\left.
\left(\Lambda\frac{\partial}{\partial \Lambda}\Lambda\right)^{n} f(p) =
\sum_{k=1} \Lambda^{k}\widetilde { W}_{k -n}^{(-,n)}(p)
f(p)
\right|_{{p_a =  { \Tr} \Lambda^{a}}}
\label{W-tilde-2}
\ee
Now one can use the manifest expressions for the commutative families in $W_{1+\infty}$ algebra with $m=\pm 1$, \cite[Eqs.(51)-(54)]{MMMP1} and note that
\be
\hat H^{(1)}_n=\Tr \Big(\Lambda{\p\over\p\Lambda}\Lambda\Big)^n=\sum_{k\ge 1}p_k\widetilde { W}_{k -n}^{(-,n)}(p)\nn\\
\hat H^{(-1)}_{-n} =\Tr \Big({\p\over\p\Lambda}\Big)^n=\sum_{k\ge 1}p_k\widetilde { W}_{k + n}^{(+,n)}(p)
+N\widetilde { W}_{n}^{(+,n)}(p)
\ee
This series of commuting Hamiltonians is associated with the rational Calogero model at the free fermion point \cite{MMCal}.

Another possibility of defining the $\widetilde W$ generators is to use the recurrent relations that follow from \eqref{W-tilde-1}-\eqref{W-tilde-2}:
\be
\widetilde{W}_{k}^{(-,n+1)}(p) = \sum_{a\geq 1} p_a
\widetilde{W}_{k + a}^{(-,n)}(p) +
\sum_{a=1}^{k+n-1}a\frac{\partial}{\partial p_a}
\widetilde{W}_{k-a}^{(-,n)}(p)+N\widetilde{W}_{k}^{(-,n)}
\label{W-tilde-rec1}\\
\widetilde{W}_{k }^{(+,n+1)}(p) = \sum_{a\geq 1} p_a
\widetilde{W}_{k + a}^{(+,n)}(p) +
\sum_{a=1}^{k-n}a\frac{\partial}{\partial p_a}
\widetilde{W}_{k-a }^{(+,n)}(p)+N\widetilde{W}_{k}^{(+,n)}
\label{W-tilde-rec13}
\ee
supplemented with ``the initial condition''
\be\label{WI1}
\begin{array}{c}
\widetilde{W}^{(\ast,1)}_k = k\frac{\partial}{\partial p_k},\ \ k\geq 1\cr
\cr
\widetilde{W}^{(\ast,1)}_0 = N\cr
\end{array}
\ee
and one generally requires that
\be\label{rec}
\widetilde{W}^{(-,n)}_k = 0,\ \ k\leq -n\nn\\
\widetilde{W}^{(+,n)}_k = 0,\ \ k\leq n-1
\ee
We do not write down here commutation relations of the $\widetilde W$ algebras, they can be found in \cite{Gava,GKMU,UFN}.
In order to improve the notation, from now on, we denote $\widetilde{W}_{k}^{(\mp 1,n)}$ just as $W^{(\pm 1,\pm n)}_k$ and, more generally, $W^{(m,n)}_k$ keeping in mind that $m$ may be both positive and negative (and zero).

\section{$W_{1+\infty}$ and $\widetilde W$ algebras}

Similarly to formulas (\ref{W-tilde-1})-(\ref{W-tilde-2}) of the previous section, here we introduce an extension of the $\widetilde W$ algebras via the definitions
\be
\boxed{
\left.
 \Big(\Lambda^{-1}\hat {\cal D}^m\Big)^n f(p) =
\sum_{k=-\infty}^\infty \Lambda^{k}\widetilde { W}_{k + n}^{(-m,-n)}(p)
f(p)
\right|_{{p_a =  { \Tr} \Lambda^{a}}}
}
\label{W-tilde-mn-1}
\ee
and
\be
\boxed{
\left.
\Big(\hat{\cal D}^m\Lambda\Big)^n f(p) =
\sum_{k=-\infty}^\infty \Lambda^{k}\widetilde { W}_{k -n}^{(m,n)}(p)
f(p)
\right|_{{p_a =  { \Tr} \Lambda^{a}}}
}
\label{W-tilde-mn-2}
\ee
where $\hat{\cal D}=\Lambda\frac{\p}{\p \Lambda}$. These are the generalized $\widetilde W$ algebras that we study in this paper.

Notice that the sums over $k$ in (\ref{W-tilde-1}) and (\ref{W-tilde-2}) run over all integers, while, comparing with the l.h.s., only terms with $k\ge 0$ contribute (in fact, in (\ref{W-tilde-2}) even with $k>0$). This guarantees the relation similar to (\ref{rec}):
\be
\widetilde{W}^{(m,n)}_k = 0,\ \ k\le -n-{\cal H}(-n)
\ee
where ${\cal H}(n)$ is the Heaviside function.

Using this definition and \cite[Eqs.(51)-(54)]{MMMP1}, one immediately obtains the relations between the commutative families of Hamiltonians of the $W_{1+\infty}$ algebra and $\widetilde W$-generators:
\be\label{EF}
\hat H^{(m)}_n=\Tr \Big(\hat{\cal D}^m\Lambda\Big)^n=\sum_{k\ge 1}p_k\widetilde { W}_{k -n}^{(m,n)}(p)\nn\\
\hat H^{(-m)}_{-n} =\Tr  \Big(\Lambda^{-1}\hat {\cal D}^m\Big)^n=\sum_{k\ge 1}p_k\widetilde { W}_{k + n}^{(-m,-n)}(p)
+N\widetilde { W}_{n}^{(-m,-n)}(p)
\ee
In fact, these two formulas can be written in the universal form describing the relation of the $\widetilde W$ algebras with the commuting Hamiltonians in $W_{1+\infty}$:
\be\label{WinW}
\boxed{
\hat H^{(m)}_n=\sum_{k\ge 0}p_k\widetilde { W}_{k-n}^{(m,n)}(p)
}
\ee
where we put $p_0=N$.

\section{A recursive definition of the generalized $\widetilde W$ algebra}

Now we are going to use a counterpart of the recurrent relations (\ref{W-tilde-rec1}), (\ref{W-tilde-rec13}) 
in order to find an equivalent definition of the $\widetilde W$ generators. In the case of $m\ne 1$, the relations become more involved, in particular, they require to use at intermediate stages some more elements of the $W_{1+\infty}$ algebra, which do not belong to commutative families.

Thus, let us define an extended algebra
\be
\left.
 \Lambda^{-1}\hat {\cal D}^l\Big(\Lambda^{-1}\hat {\cal D}^m\Big)^{n - 1} f(p) =
\sum_{k=-\infty}^\infty \Lambda^{k}\widetilde { W}_{k + n}^{(-m,-n|-l)}(p)
f(p)
\right|_{{p_a =  { \Tr} \Lambda^{a}}}
\label{W-tilde-mn-ext-1}
\ee
and
\be
\left.
\hat{\cal D}^l\Lambda\Big(\hat{\cal D}^m\Lambda\Big)^{n - 1} f(p) =
\sum_{k=-\infty}^\infty \Lambda^{k}\widetilde { W}_{k -n}^{(m,n|l)}(p)
f(p)
\right|_{{p_a =  { \Tr} \Lambda^{a}}}
\label{W-tilde-mn-ext-2}
\ee
so that
\be\label{WW}
\widetilde { W}_{k }^{(m,n|m)}(p)=\widetilde { W}_{k}^{(m,n)}(p)
\ee
and
\be
\widetilde{W}^{(\ast, n|\ast )}_k = 0, \qquad k\le -n-{\cal H}(-n)
\ee
Then, there are two types of recurrent relations: there are relations that allow one to express $\widetilde { W}_{k }^{(m, \pm n | \pm l)}(p)$ from $\widetilde { W}_{k }^{(m, \pm n|\pm 1)}(p)$:
\be\label{rr1-}
\boxed{
\widetilde{W}_{k}^{(-m, -n | -l - 1)} = \sum_{r \ge 1}^{} p_r \widetilde{W}_{k + r}^{(-m, -n | -l)} + \sum_{r = 1}^{k - n}r \frac{\partial }{\partial p_r}  \widetilde{W}_{k - r}^{(-m, -n | -l)} + N \widetilde{W}_{k}^{(-m, -n | -l)},}
\ee
\be\label{rr1+}
\boxed{
\widetilde{W}_{k}^{(m, n | l + 1)} = \sum_{r \ge 1}^{} p_r \widetilde{W}_{k + r}^{(m, n | l)}
+ \sum_{r = 1}^{k + n - 1} r \frac{\partial }{\partial p_r}\widetilde{W}_{k - r}^{(m, n | l)} + N
\widetilde{W}_{k}^{(m, n | l)}.}
\ee
i.e. using these relations, one can construct $\widetilde { W}_{k }^{(m,n|m)}(p)$, i.e. $\widetilde { W}_{k}^{(m,n)}(p)$ from $\widetilde { W}_{k }^{(*,\pm 1|\pm 1)}(p)$.
There is another type of relations:
\be\label{rr2-}
\boxed{
\widetilde{W}_{k}^{(-m, - n - 1 | -1)} = \sum_{r \ge 1}^{} p_r \widetilde{W}_{k + r}^{(-m, -n)} + \sum_{r = 1}^{k - n}r \frac{\partial }{\partial p_r}  \widetilde{W}_{k - r}^{(-m, -n)} + N \widetilde{W}_{k}^{(-m, -n)}
,}
\ee
\begin{equation}\label{rr2+}
\boxed{\widetilde{W}_{k}^{(m,  n + 1 | 1)} = \sum_{r \ge 1}^{}p_r \widetilde{W}_{k + r}^{(m, n)}
 + \sum_{r = 1}^{k +n - 1} r \frac{\partial }{\partial p_r}  \widetilde{W}_{k - r}^{(m, n)} +N \widetilde{W}_k^{(m, n)} ,}
\end{equation}
supplemented by the initial conditions as in (\ref{WI1}):
\be\label{ic}
\begin{array}{c}
\widetilde{W}^{(\ast,\pm 1|\pm 1)}_{k} = k\frac{\partial}{\partial p_k},\ \ k\geq 1,\cr
\cr
\widetilde{W}^{(\ast, 1| 1)}_0 = N,\cr
\end{array}
\ee
Using these relations, one can start from $(m,n,l)=(m,1,1)$ and use (\ref{ic}), then, raise $l=1$ up to $m$ using (\ref{rr1+}), and use (\ref{WW}) in order to obtain $\widetilde W^{(m,1)}$. Then, one uses (\ref{rr2+}) in order to obtain $(n,l)=(2,1)$, etc. A similar procedure is also applied for negative $m$ and $n$.
Thus, one obtains with these recurrent relations all $\widetilde W^{(m,n)}$ generators.

Note that sometimes it possible to find direct recurrent relations involving only $\widetilde { W}_{k}^{(m,n)}$ generators. For instance,
generators of the algebra $\widetilde { W}_{k}^{(m,1)}$ satisfy the relations
\be
\widetilde{W}_{k }^{(m+1,1)}(p) = \sum_{a\geq 1} p_a
\widetilde{W}_{k + a}^{(m,1)}(p) +
\sum_{a=1}^{k}a\frac{\partial}{\partial p_a}
\widetilde{W}_{k-a }^{(m,1)}(p)+N\widetilde{W}_{k}^{(m,1)}
\label{W-tilde-rec2}
\ee
with the same ``initial condition" as before
\be\label{WI2}
\begin{array}{c}
\widetilde{W}^{(1,1)}_k = k\frac{\partial}{\partial p_k},\ \ k\geq 1\cr
\cr
\widetilde{W}^{(1,1)}_0 = N\cr
\end{array}
\ee

A detailed description of these recurrent relations as well as manifest examples of the generators evaluated using this procedure can be found elsewhere \cite{Dr}.

\section{$\widetilde W^{(0,n)}$ algebras, completed cycles and $\tau$-functions}

So far we considered the algebras with positive and negative $m$, however, one can also extend these relations to generators of the $\widetilde W^{(0,n)}$ algebras. These algebras are associated with the $W_{1+\infty}$ commutative family corresponding to the vertical ray in Fig.1, the operators of this family
$\hat{\cal H}_{i+j}=[\hat F_i,\hat E_j]=[\hat H_1^{(i)},\hat H_{-1}^{(-j)}]$ can be obtained in the matrix realization from formulas (\ref{EF}):
\be\label{calH}
\hat{\cal H}_{2}= 2\Tr\hat{\cal D}+N^2\nn\\
\hat{\cal H}_{3}=3\Tr\hat{\cal D}^2 +3N\Tr\hat{\cal D}+N^3= 6\hat W_0\nn\\
\hat{\cal H}_{4}=4\Tr\hat{\cal D}^3+4N\Tr\hat{\cal D}^2+4N^2\Tr{\cal D}+N^4+2\left(\Tr\hat{\cal D}\right)^2\nn\\
\ldots\nn\\
\hat{\cal H}_{n+1}=\sum_{j=0}^{n}\widehat\Tr\hat{\cal D}^j\Big(\hat{\cal D}+I\widehat\Tr\Big)^{n-j}\cdot I
\ee
where $I$ is the unit matrix and the operator $\widehat\Tr$ acts\footnote{For instance,
$$
\widehat\Tr \Big(\hat{\cal D}+I\widehat\Tr\Big)^2\cdot I=\widehat\Tr\Big(\hat{\cal D}^2+I\widehat\Tr\hat{\cal D}+\hat{\cal D}\widehat\Tr+I\widehat\Tr\cdot I\widehat\Tr\Big)
\cdot I=\Tr\hat{\cal D}^2+N\Tr\hat{\cal D}+N\Tr\hat{\cal D}+N^3
$$
}
just by producing the trace, in particular, $\widehat\Tr\cdot I=N$.

These generators are nothing but the commutative Hamiltonians of the trigonometric Calogero-Sutherland model at the free fermion point \cite{MMCal}. In particular, the operator $\hat W_0$ is the cut-and-join operator \cite{GJ,MMN} which is known to be the trigonometric Calogero-Sutherland Hamiltonian \cite{MS}.

It should not come as a surprise that the operators $\hat {\cal H}_n$ are not expressed through single trace operators. The reason is as follows. Consider the partition function generated by such an operator:
\be
{\cal Z}_n=e^{{1\over n}\hat{\cal H}_n}\cdot e^{\sum_{k=1}{1\over k}g_kp_k}
\ee
This partition function can be presented in the form
\be\label{cZn}
{\cal Z}_n=\sum_R S_R\{g_k\}S_R\{p_k\}e^{C_n(R)}
\ee
where $C_n(R)$ is the eigenvalue of an $n$-th Casimir operator associated with $\hat{\cal H}_n$. As soon as $\hat{\cal H}_n$ is an element of the $W_{1+\infty}$ algebra, (\ref{cZn}) is a KP $\tau$-function \cite{Orlov,TT}. However, the sum over partitions (\ref{cZn}) is a KP $\tau$-function iff $C_n(R)$ is a linear combination of quantities $\Lambda_R:=\sum_i(R_i-i+1/2)^k-(-i+1/2)^k$ \cite{GKM2,OS,Oko,OkOl,Lala,AMMN1}. Such $\tau$-functions are called hypergeometric \cite{OS}. In fact, in such a case, it is a KP $\tau$-function w.r.t. the both sets of times, $p_k$ and $g_k$ , and the dependence on the Toda zeroth (discrete) time requires further specification  (see \cite{Taka}) giving rise to the Toda lattice hierarchy \cite{UT}.

Now let us note that the basis in the space of single trace operators $\Tr\hat{\cal D}^k$ is provided by the generalized cut-and-join operators $\hat W_{[k]}$ \cite{MMN}\footnote{For instance, $$\hat W_0=\hat W_{[2]}-N\hat W_{[1]}-{N^3\over 6}$$}, the Schur functions being eigenfunction of these operators:
\be
\hat W_{[k]}\ S_R\{p_k\}=\phi_R([k])\ S_R\{p_k\}
\ee
where the eigenvalue $\phi_R([k])$ is proportional \cite{IK,MMN,MMN2}\footnote{An explicit formula for $\phi_R([k])$ through the shifted Schur functions \cite{OK} can be found, e.g., in \cite{MMN3}.} to the value of character of the symmetric group ${\cal S}_n$, $n=|R|$ in the representation $R$ on the element with the only non-unit cycle of length $k$ \cite{Fulton}. The action of operator $\hat W_{[k]}$ is
\be
e^{\hat W_{[k]}}\cdot 1=\sum_R S_R\{g_k\}S_R\{p_k\}e^{\phi_R([k])}
\ee
because of the Cauchy identity
\be
e^{\sum_{k=1}{1\over k}g_kp_k}=\sum_R S_R\{g_k\}S_R\{p_k\}
\ee
Thus, since ${\cal Z}_n$, (\ref{cZn}) should be a hypergeometric $\tau$-function,  the Schur function $S_R\{p_k\}$ should be an eigenfunction of $\hat{\cal H}_n$ with an eigenvalue being a linear combination of the quantities $\Lambda_R$.
The point is, however, that $\phi_R([k])$ is not such a linear combination at $k>2$ \cite{MMN}, and only a non-linear polynomial of $\phi_R([k])$'s is \cite{MMZh}. Such a polynomial is called \textit{completed cycle} \cite{OkOl,Lala}. This is what we observe in formulas (\ref{calH}). Completed cycles have attracted a lot of attention during the last years
in the enumerative geometry context (see, for instance, \cite{Chio08,ACEH18b,ALS16}).
In particular, they feature in the
celebrated Zvonkine's conjecture \cite{ZvoConj}, only recently proved in \cite{ZvoProof}.
With the general $\tilde{W}$-algebra point of view, advocated in the present paper,
one may naturally wonder whether quantities, built from non-vertical families
of $\tilde{W}$-operators carry equally deep enumerative geometry meaning.

In order to be more concrete, the manifest action of $\hat{\cal H}_n$ on the Schur functions, indeed, gives rise to a linear combination of $\Lambda_k$. For the sake of brevity, we choose another basis 
\be
\widetilde\Lambda_k:=\sum_i(R_i-i+1/2+N)^k-(-i+1/2+N)^k-N^k
\ee 
linearly related with the basis of $\Lambda_k$. Then,
\be
\hat{\cal H}_n\ S_R\{p_k\}={\cal E}_n(R)\ S_R\{p_k\}
\ee
with
\be
{\cal E}_{n}(R)=\sum_{j=0}{1\over 4^j}\binom{n}{2j+1}\widetilde\Lambda_{n-2j-1}+N^n
\ee

Now let us note that formula (\ref{calH}) can be rewritten in the form
\be
\hat{\cal H}_{n+1}=\sum_{j=0}^n\Tr\hat{\cal D}^j\left(\Lambda^{-1}\hat{\cal D}\Lambda\right)^{n-j}
\ee
This paves a way for introducing the generators of the $\widetilde W^{(0,n)}$ algebra either from the relation
\be\label{W0}
\boxed{
\left.\sum_{j=0}^n\hat{\cal D}^j\left(\Lambda^{-1}\hat{\cal D}\Lambda\right)^{n-j}f(p) =
\sum_{k=-\infty}^\infty \Lambda^{k}\widetilde { W}_{k}^{(0,n)}(p)
f(p)
\right|_{{p_a =  { \Tr} \Lambda^{a}}}}
\ee
or from the recurrent relations that follow from (\ref{W0}). To this end, we again need to introduce auxiliary operators
\be
\left.\left(\Lambda^{-1}\hat{\cal D}\Lambda\right)^{n}f(p) =
\sum_{k=-\infty}^\infty \Lambda^{k}\widetilde { W}_{k}^{(0,n|l)}(p)
f(p)
\right|_{{p_a =  { \Tr} \Lambda^{a}}}, \qquad n < l
\ee
\be
\left.\hat{\cal D}^{n-l}\left(\Lambda^{-1}\hat{\cal D}\Lambda\right)^{l}f(p) =
\sum_{k=-\infty}^\infty \Lambda^{k}\widetilde { W}_{k}^{(0,n|l)}(p)
f(p)
\right|_{{p_a =  { \Tr} \Lambda^{a}}}, \qquad n \ge l
\ee
which satisfy the recurrent relations
\be\label{rr0}
\widetilde{W}_k^{(0, n + 1| l)} = \sum_{r \ge 1}^{} p_{r }\widetilde{W}_{k + r}^{(0, n| l)} + \sum_{r = 1}^{k} r \frac{\partial }{\partial p_r}  \widetilde{W}_{k - r}^{(0, n| l)}  + N \widetilde{W}_{k}^{(0, n| l)}
\ee
along with the initial conditions:
\be\label{ic0}
\widetilde{W}_k^{(0, 0| l)} = \delta_{k, 0}
\ee
\be
\widetilde{W}_k^{(0, n | l)} = 0 \qquad \text{for} \qquad k\le -{\cal H}(l-n)
\ee
These auxiliary generators are clearly summed into $\widetilde W^{(0,n)}_k$:
\be\label{sum0}
\widetilde W^{(0,n)}_k=\sum_{l=0}^n\widetilde W^{(0,n|l)}_k
\ee
Hence, for evaluating $\widetilde{W}_k^{(0, n)}$, one has to start from $\widetilde{W}_k^{(0, 0| l)}$ in (\ref{ic0}) and then, using (\ref{rr0}), to obtain all $\widetilde{W}_k^{(0, p| l)}$ with $p\le n$. This evaluation has to be done at each $l\le n$, and then one can use formula (\ref{sum0}) in order to finally obtain $\widetilde{W}_k^{(0, n)}$.

From relation (\ref{W0}), it immediately follows that 
\be
\boxed{
\hat{\cal H}_n=\sum_{k\ge 0}p_k\widetilde W^{(0,n)}_k
}
\ee

\section{$\widetilde W^{(k,m)}$ algebras as Ward identities in the WLZZ matrix models}

After having constructed the generalized $\widetilde W$ algebras, we are ready to discuss the models where they form algebras of constraints. The basic example is given by the two-matrix model:
\be\label{im}
Z_{n}=\int\int_{N\times N}dXdY\exp\left(-\Tr XY+\sum_k {p_k\over k}\Tr X^k+{1\over n}\Tr Y^n\right)
\ee
where the integral is understood as integration of a power series in $p_k$, and $X$ are Hermitian matrices, while $Y$ are anti-Hermitian ones. This matrix integral at $n>1$ satisfies a set of the $\widetilde W^{(1,n)}$ algebra constraints \cite{AMMN2,MMsc},
\be
\widetilde {W}_{k}^{(1,n)}Z_{n}=(n+k){\p Z_{n}\over\p p_{n+k}}, \ \ \ \ \ k\ge -n+1
\ee
At the same time, one can follow paper \cite{MMMR} in order to encode all these constraints in a single equation,
\be
\Bigg(\sum_{a=1}ap_a{\p\over\p p_a}-\underbrace{\sum_{k=0}p_k\widetilde {W}_{k-n}^{(1,n)}}_{\hat H_n^{(1)}}\Bigg)
Z_n=0
\ee
its solution being
\be
Z_n=e^{{1\over n}\hat H_n^{(1)}}\cdot 1
\ee
and, hence, one associates $Z_n=Z^{(1)}_n$.

This scheme is completely extended to the whole series $Z_n^{(m)}$ (though it is no longer a matrix integral): the partition function
\be
Z_n^{(m)}=\sum_R
\left({S_R\{p_k=N\}\over S_R\{p_k=\delta_{k,1}\}}\right)^mS_R\{p_k=\delta_{k,n}\}S_R\{p_k\}
\ee
satisfies the Ward identities
\be
\widetilde {W}_{k}^{(m,n)}Z_{n}^{(m)}=(n+k){\p Z_{n}^{(m)}\over\p p_{n+k}}, \ \ \ \ \ k\ge -n+1
\ee
or the single equation
\be
\Bigg(\sum_{a=1}ap_a{\p\over\p p_a}-\underbrace{\sum_{k=0}p_k\widetilde {W}_{k-n}^{(m,n)}}_{\hat H_n^{(m)}}\Bigg)
Z_n^{(m)}=0
\ee
so that
\be\label{Znm}
Z_n^{(m)}=e^{{1\over n}\hat H_n^{(m)}}\cdot 1
\ee
in accordance with \cite{Ch1,Ch2}. This set of partition functions $Z_n^{(m)}$ was first introduced in \cite{China1,China2} (in the case of $m=1$), hence the name WLZZ models, and was later extended to arbitrary integer $m$ in \cite{Ch1,Ch2}. Note also that the set of partition functions associated with $\widetilde W^{(0,n)}$ was considered in \cite{AMMN2}
(along with its matrix model realization, see \cite{AMMN2}, it was called there $Z_{(1,m)}$).

As for the series of  $\widetilde W^{(m,n)}$ algebras with negative $m$, they generate the partition functions \cite{Ch1,Ch2}
\be\label{Zneg}
Z_{-n}^{(-m)}=e^{{1\over n}\hat H_{-n}^{(-m)}}\cdot e^{\sum_{k=1}{1\over k}g_kp_k}
\ee
where $g_k$ are non-zero parameters, since action on unity would give a trivial answer. Hence, the Hamiltonians $H_{-n}^{(-m)}$ do not give rise to a single equation and are not related to a constraint algebra, and neither are the corresponding $\widetilde W^{(-m,-n)}$.

Similarly, one can generate the Hurwitz partition functions corresponding to the completed cycles (see a discussion in \cite{MMsc,Poptr}),
\be\label{Z0}
Z_{n}^{(0)}=e^{{1\over n}\hat{\cal H}_{n}}\cdot e^{\sum_{k=1}{1\over k}g_kp_k}
\ee
In the both these cases the algebra of constraints has to be constructed yet. 

\section{Concluding remarks}

In this paper, extending earlier known $\widetilde W^{(\pm,n)}$ algebras, we constructed a series of generalized $\widetilde W^{(m,n)}$ algebras labelled by two integer numbers $m$ and $n$ that are either both negative, or both are non-negative. These algebras are related to commutative subalgebras of the $W_{1+\infty}$ algebra associated with integer rays \cite{MMMP1}. In fact, each element of such a subalgebra $\hat H^{(m)}_n$ is given by a simple formula connecting it with a $\widetilde W^{(m,n)}$ algebra, (\ref{WinW}).

We presented the definition of the generalized $\widetilde W$ algebra as an algebra of differential operators in terms of variables $p_k$ both via a formulation in terms of matrix derivatives, and via a recursive definition. This allows one to construct the $\widetilde W$ operators manifestly. Note that explicit formulas for $\hat H^{(m)}_n$ in terms of variables $p_k$ were recently presented in \cite{MMMP1}. Here we provide an alternative set of formulas for $\hat H^{(m)}_n$, which is based on the manifestly constructed $\widetilde W$ operators and formula (\ref{WinW}).

Note that the recursive definition of the $\widetilde W$ algebra requires an auxiliary set of operators from $W_{1+\infty}$, which do not belong to commutative families. The basic role of these operators, however, remains unclear, and we postpone studying these operators to further studies.

The $\widetilde W$ algebras have originally appeared as algebras of constraints in matrix models. The partition functions generated by $\hat H^{(m)}_n$ with positive $m$ and $n$ as the operators determining the $W$-representation of matrix models, (\ref{Znm}) are called the WLZZ models \cite{China1,China2,Ch1,Ch2}, and we demonstrated that these partition functions are satisfied by the set of constraints given by the generalized $\widetilde W$ algebras. Unfortunately, the algebra of constraints for the partition functions generated by the operators $\hat H^{(m)}_n$ with non-positive $m$, (\ref{Zneg}), (\ref{Z0}) is not described yet, only the case of $m=-1$, $n=-2$ was studied in
\cite[see sec.5.2 and especially formula (104)]{MMsc}, where it was demonstrated that, even in this simplest case, the algebra of constraints is given by linear combinations of generators of algebras $\widetilde W^{(-1,n)}$ with {\it all} negative $n$. The problem of finding algebras of constraints for the partition function $Z_n^{(m)}$ with arbitrary negative $m$ and $n$ also deserves further investigation.

Another important issue that was not touched in the present paper is a $\beta$-deformation of the $\widetilde W$ algebras. Such a deformation is definitely possible, since, as we demonstrated in \cite{MMMP2}, the commutative families associated with the integer rays of the $W_{1+\infty}$ algebra are immediately lifted to the affine Yangian algebra, which exactly provides the required $\beta$-deformation. We are planning to return to this issue elsewhere.

\section*{Acknowledgements}

We are grateful to V. Mishnyakov, A. Morozov and M. Tsarkov for useful discussions.
This work was partly supported by grants RFBR 21-51-46010-ST-a and 21-52-52004-MNT-a, and by the grants of the
Foundation for the Advancement of Theoretical Physics and Mathematics ``BASIS".

\end{document}